\documentclass[conference]{IEEEtran}
\IEEEoverridecommandlockouts

\usepackage{cite}
\usepackage{amsmath,amssymb,amsfonts}
\usepackage{subcaption}
\usepackage{algpseudocode}
\usepackage{graphicx}
\graphicspath{ {./images/} }
\usepackage{textcomp}
\usepackage{xcolor}
\usepackage{makecell}
\usepackage[linesnumbered,ruled,norelsize]{algorithm2e}
\makeatletter
\newcommand{\removelatexerror}{\let\@latex@error\@gobble}
\makeatother
\usepackage{multicol}
\usepackage{mwe}
\usepackage{graphicx}
\usepackage{epsfig}
\usepackage{epstopdf}

\SetCommentSty{mycommfont}

\SetKwInput{KwInput}{Input}                
\SetKwInput{KwOutput}{Output}              

\def\BibTeX{{\rm B\kern-.05em{\sc i\kern-.025em b}\kern-.08em
    T\kern-.1667em\lower.7ex\hbox{E}\kern-.125emX}}

\begin{document}

\title{Energy-saving Cross-layer Optimization of Big Data Transfer Based on Historical Log Analysis\\
}

\author{\IEEEauthorblockN{Lavone Rodolph, MD S Q Zulkar Nine, Luigi Di Tacchio, and Tevfik Kosar} \IEEEauthorblockA{Department of Computer Science and Engineering, University at Buffalo, Buffalo, New York 14260, USA} \IEEEauthorblockA{
Email: \{lrodolph, mdsqzulk, luigidit,  tkosar\}@buffalo.edu}
}

\maketitle
\begin{abstract}
With the proliferation of data movement across the Internet, global data traffic per year has already exceeded the Zettabyte scale. The network infrastructure and end-systems facilitating the vast data movement consume an extensive amount of electricity, measured in terawatt-hours per year. This massive energy footprint costs the world economy billions of dollars partially due to energy consumed at the network end-systems. Although extensive research has been done on managing power consumption within the core networking infrastructure, there is little research on reducing the power consumption at the end-systems during active data transfers. This paper presents a novel cross-layer optimization framework, called Cross-LayerHLA, to minimize energy consumption at the end-systems by applying machine learning techniques to historical transfer logs and extracting the hidden relationships between different parameters affecting both the performance and resource utilization. It utilizes offline analysis to improve online learning and dynamic tuning of application-level and kernel-level parameters with minimal overhead. This approach minimizes end-system energy consumption and maximizes data transfer throughput. Our experimental results show that Cross-LayerHLA outperforms other state-of-the-art solutions in this area. 
\end{abstract}


\section{Introduction}
With the vast amount of data produced by scientific and engineering applications, social media, cloud computing, and the emerging Internet of Things (IoT), it is estimated that by 2022 the global Internet traffic will exceed 4.8 zettabytes per year caused by a projected 28.5 billion network-connected devices~\cite{CiscoWhitePaper2020}. 
The energy consumption of telecommunication networks has already exceeded 350 terawatt-hours, and the Internet comprises more than 10\% of the overall energy consumption in many countries, costing the global economy billions of dollars per year~\cite{GWATT}.
This massive energy footprint has ignited immense research in power-aware networking, focusing on reducing power consumption at both the hardware and systems-levels, including network devices.

A large portion of the existing energy-efficient networking research focuses on reducing power consumption within the core networking infrastructure (e.g., switches, hubs, and routers). State-of-the-art power-aware networking techniques include emerging architectures with programmable switches~\cite{greenberg2008towards}, power-aware networking protocols designed to consider energy consumption while routing data~\cite{chabarek2008power}, and putting idle components to sleep~\cite{gupta2003greening}. On the other hand, many of the existing core infrastructure solutions have shortcomings. For example, putting idle components to sleep can be detrimental to throughput. Also, replacing existing network switches and network protocols with power-aware switches and energy-efficient network protocols are expensive solutions and not practical in the short term. This paper presents a cost-effective, energy-efficient, practical end-system solution, called Cross-LayerHLA, which is an optimization framework that combines offline historical log analysis with dynamic online tuning to minimize end-system energy consumption and maximize data transfer throughput.

The major contributions of Cross-LayerHLA include: 
\begin{itemize}
    \item To the best of our knowledge, it is the first to integrate cross-layer adaptive online tuning with offline analysis using machine learning techniques to reduce energy consumption at the end systems and improve data transfer throughput performance.
    \item It is the first to dynamically tune both application-level and kernel-level data transfer parameters while obtaining optimal parameter values from offline analysis based on real-time network conditions.
    \item Compared to state-of-the-art solutions, it reduces end-system energy consumption during data transfers considerably while increasing the data transfer throughput at the same time.
\end{itemize}
 
The rest of this paper is organized as follows: Section \textrm{II} describes the related work in the field; 
Section \textrm{III} describes how we model and optimize energy consumption and throughput performance based on historical log analysis and dynamic online tuning;   Section \textrm{IV} evaluates Cross-LayerHLA and compares it to other state-of-the-art solutions in this area;  and Section \textrm{V} concludes the paper. 

\section{Related Work}

Prior work on application level tuning of transfer parameters mostly proposed static or non-scalable solutions to the problem with some predefined values for the subset of the problem space~\cite{R_Hacker02, R_Dinda05, deelman2006makes, kola2004disc, kosar2011stork}. The main problem with such solutions is that they do not consider the dynamic nature of the network links and the background traffic in the intermediate nodes. There has been several efforts in 
Kasparan et al.~\cite{kaspar2010using} analyzed how pipelining affects throughput in local area networks and high-speed downlink packet access networks. Kim et al.~\cite{kim2015highly} and Yildirim et al.~\cite{yildirim2012gridftp} considered the combined effect of parallelism, pipelining, and concurrency on end-to-end data transfer throughput.   
Natarajan et al.~\cite{natarajan2009multiple} showed that using a single stream for transferring independent web objects is very inefficient in high latency networks. Using concurrent channels for delivering such objects increases download rates and decreases browser response times by enabling concurrent rendering. 
Robert et al.~\cite{kuschnig2010improving} used parallel file requests to increase usage of available bandwidth during Internet video streaming. 
Li et al.~\cite{liu2011parallel} presented a parallel streaming method instead of traditional adaptive sequential streaming. In distributed network environments, fetching media segments by parallel channels increases streaming performance and also copes with inefficient usage of resources.
Alan et al.~\cite{alan2015energy} analyzed the effects of different application-layer protocol parameters such as TCP pipelining, parallelism and concurrency levels on end-to-end throughput versus total energy consumption.

In our prior work~\cite{di2019cross}, we combined application-level (pipelining, parallelism, and concurrency) and kernel-level parameters (number of active cores, CPU frequency level) to minimize energy consumption at the end systems (source and destination nodes) during active data transfers. In that work, we heuristically performed the optimization without considering historical log analysis, but its convergence to optimal values was slow. In this work, we optimize data transfers and reduce energy consumption by jointly employing offline historical analysis, machine learning techniques, and online dynamic real-time tuning.

\section{Modeling and Optimizing Transfer Throughput and Energy Consumption}
To accurately model the data transfer throughput and end-system energy consumption, we consider and ensure that:
(1) our model is representative and characteristic of real-world data transfers; (2) our model accounts for fluctuating networking conditions, external background traffic, and external network loads; and (3) our model considers dataset characteristics (dataset size, average file size within a dataset, and file size distribution). The dataset characteristics would have a significant impact on transfer throughput performance and end-system energy consumption~\cite{alan2015energy}. For example, transferring a dataset comprised of multiple small files (such as text files) can benefit significantly from high concurrency and pipelining, and transferring a single large file would benefit from parallelism. To convey the impact of the kernel-level and application-level transfer parameters on end-system energy consumption and throughput performance, we express the relationships as the following functions:  
\begin{equation}
\label{Energy_Equation}
E = f_1(cpu_{num}, cpu_{freq}, cc, p, pp, data, net)\newline 
\end{equation}
\begin{equation}
\label{Throughput_Equation}
T = f_2(cpu_{num}, cpu_{freq}, cc, p, pp, data, net) 
\end{equation}
where $E$ is the energy consumption, $T$ is the throughput performance, $cpu_{num}$ is the number of active CPU cores, $cpu_{freq}$ is the frequency level of each active CPU core, $cc$ is the concurrency level, $p$ is the parallelism level, $pp$ is the pipelining level, $data$ is the dataset characteristics and $net$ is the network characteristics.

To address the factors above, we chose to: 
(1) accumulate and store real-world historical log data transfer information; 
(2) utilize machine learning to perform clustering on the historical log files, grouping/stratifying similar log entries based on network conditions, dataset meta-information, and network characteristics to uncover hidden relationships; and (3) derive an appropriate interpolation or regression model based on the rendered data characteristics. These steps are explained in the offline optimization subsection below.

\subsection{Offline Analysis}

\textbf{Step 1: Accumulate and cache historical log data:}
During a data transfer, we accumulate the following information: i) throughput performance; ii) energy consumption; iii) network characteristics; iv) end-system resource usage; v) dataset characteristics; and vi) kernel-level and application-level parameter configuration.
Kernel-level and application-level parameter configuration includes: i) the number of active CPU cores; ii) the frequency level in which the active CPU cores operate; iii) concurrency; iv) parallelism; and v) pipelining. Historical log files are accumulated periodically and cached in a log server. 

\textbf{Step 2: Stratify historical log data:}
After accumulating historical log data, we stratify the log entries based on similar data characteristics. We utilize Hierarchical Agglomerative Clustering (HAC)~\cite{murtagh2014ward} with the Unweighted Pair Group Method with Arithmetic Mean (UPGMA)~\cite{gronau2007optimal} to achieve this. Also, we enforce artificial boundaries based on data transfers' distribution characteristics to ensure accurate stratification. Boundaries are based on the mean, and standard deviation of the achieved throughput and energy consumption of multiple collective data transfer runs. 
We utilize a bottom-up approach to cluster data in tiers. For tier 1, we cluster log entries based on external network characteristics. For tier 2, we apply agglomerative hierarchical clustering within each of the tier 1 clusters, further grouping log entries based on dataset meta-info (dataset size, file distribution, average file size and number of files in dataset). Since high external network loads decreases throughput performance and increases end-system energy consumption we further stratify the strata/data into interval ranges based on the data transfer logs' external network load distribution characteristics. This is necessary as current optimal parameters may become suboptimal as external network traffic changes. 
For tier 3, we further group tier 2 clusters based on network characteristics such as source and destination nodes and their corresponding bandwidth and latency.

\textbf{Step 3: Derive interpolation/regression model:}
After data is accumulated, stratified, and categorized, we analyze and extract the hidden relationships within them. Since estimated energy consumption and throughput can be modeled as polynomial surfaces, we utilized piece-wise cubic spline interpolation ~\cite{ bickley1968piecewise} to derive the interpolants for each cluster/stratum. Second-order (quadratic) polynomial interpolation and third-order (cubic) polynomial interpolation based on Newton's or Lagrange's methods ~\cite{kincaid2009numerical} do not fit the data as smoothly as piece-wise cubic polynomials, which assures smoothness up to the second derivative. Since each data transfer parameter has distinct qualities, we model them independently. We first model a 2-dimensional cubic spline interpolation for energy consumption utilizing the number of CPU cores, denoted as cpu, as the abscissas and the end-system energy consumption values, denoted as e, as the ordinates. 

\begin{equation}
\label{Energy_Equation}
e_i = s_i(cpu_i) 
\end{equation} 

Given a cluster/stratum of discrete points which we will refer to as knots in a 2-dimensional space $\{(cpu_i,e_i)\}$, $i=1,...,N$, we formulate the interpolant $e_i = s(cpu_i)$ by using the piece-wise polynomial $s_i(cpu)$ as a nexus bridging together the pair of consecutive knots $(cpu_i,e_i)$ and $(cpu_{i+1},e_{i+1})$. This allows us to derive natural relaxed piece-wise cubic polynomials, where each $e_i$ have zero second derivatives at the end knots and assures curvature smoothness by restricting its coefficients. We can define each cubic polynomial piece as: 
\begin{equation}
\label{cubicPieceWisePolynomial}
s_i(cpu_i) = a_{i,0} + a_{i,1}cpu + a_{i,2}cpu^2 +  a_{i,3}cpu^3 \forall cpu \in [cpu_i, cpu_{i+1}] \newline 
\end{equation} 
Since we have $N-1$ piece-wise cubic polynomials, we have $4(N-1)$ unknown coefficients $a_{i,j}$ where $j = 0,1,2,3$ of piece-wise cubic polynomial $s_i(cpu)$. We also have $N$ continuity conditions/stipulations for all given knots evaluated by the corresponding piece-wise cubic polynomial as specified by:
\begin{equation}
\label{condition_1}
s_i(cpu_i) = e_i, \quad i = 1,..,N \newline 
\end{equation} 
Furthermore, we have $N-2$ continuity conditions/stipulations for all interior knots specified as:
\begin{equation}
\label{condition_1}
s_i(cpu_{i+1}) = s_{i+1}(cpu_{i+1}), \quad i = 1,..,N-2 \newline
\end{equation} 

Utilizing Leibniz's notation for differentiation~\cite{thurston1973leibniz}, we can specify $N-2$ slope continuity conditions/stipulations for the interior knots as:
\begin{equation}
\label{condition_1}
\frac{ds_i}{d(cpu)} (cpu_{i+1}) =  \frac{ds_{i+1}}{d(cpu)} (cpu_{i+1}) \quad i = 1,..,N-2 \newline
\end{equation} 

This produces $N-2$ quadratic polynomial continuity stipulations. We also enforce $N-2$ additional continuity stipulations for the interior knots utilizing the second derivative as follows:  
\begin{equation}
\label{condition_1}
\frac{d^2s_i}{d^2(cpu)} (cpu_{i+1}) =  \frac{d^2s_{i+1}}{d^2(cpu)} (cpu_{i+1}) \quad i = 1,..,N-2 \newline
\end{equation} 
This produces $N-2$ Linear polynomial stipulations. Since we are constructing a natural relaxed piece-wise cubic spline polynomial the second derivative at the end knots are zero and may be specified as follows:
\begin{equation}
\label{condition_1}
\frac{d^2s}{d^2(cpu)} (cpu_1) =  \frac{d^2s}{d^2(cpu)} (cpu_n) = 0 \newline
\end{equation}

By solving the system of linear equations we obtain the coefficients.\par 
End-system energy consumption also depends on the other four data transfer parameters: frequency (freq), concurrency (cc), pipelining (pp), and Parallelism (p). We extend the above example to formulate three multivariate piece-wise cubic spline polynomials corresponding to cubic polynomial surfaces. These include: $e_1=s(cpu,freq)$, $e_2=s(cc,p)$ and $e_3=s(pp)$. For each cubic spline polynomial we obtain the optimal parameter values based on data transfer service level agreements (SLA) and perform uniform sampling matching the criteria. 
Utilizing piece-wise cubic-spline surface interpolation, we were able to stitch together multiple cubic functions and predict/estimate both energy consumption and throughput performance of the previously missing parameters. This was necessary as optimal transfer parameter values needed to meet a particular SLA may not be present in the historical log data. To test the accuracy of our cubic spline interpolation method, we used 70\% of the logs to perform the interpolation and the remaining 30\% as the test data. We used the standard Root Mean Squared Error (RMSE)~\cite{wiener1949wiener} to calculate the accuracy.  

\subsection{SLA Optimization}
We developed two categories of service-level agreements (SLAs) as our optimization goals: (1) energy-constrained SLAs and (2) throughput assurance/guarantee SLAs. Based on the SLA type and the SLA specifications, we must select an optimal combination of the tunable parameters to satisfy the SLA requirements. To find the optimal parameter values for minimum end-system energy consumption, our energy-constrained optimization model determines the local minima in each of the three multivariate piece-wise cubic spline interpolation formulas and selects parameters corresponding to the minimum of the minima. This can be achieved by performing the second partial derivative test on all local minima, by calculating the Hessian matrix and determining if the matrix is positive definite. For throughput optimization, we perform the reverse, find all local maxima, calculate the Hessian matrix, and determine if the matrix is negative definite. We extend our previous optimization model ~\cite{nine2018greendataflow} to include kernel-level parameters and express it as:

\begin{equation}
\begin{aligned}
& \min_{parameters(kernel,app)} \quad & (E) \\
& \qquad \textrm{subject to} \quad  &   T \geq T_{sla} 
\end{aligned}
\end{equation}

\begin{equation}
\begin{aligned}
& \max_{parameters(kernel,app)} \quad & \int_{T_s}^{T_f} T  \\
& \qquad \textrm{subject to} \quad & E \leq E_{sla} 
\end{aligned}
\end{equation}

The energy-constrained model allows a user to select optimal data transfer parameters that maximize throughput while ensuring the achieved energy consumption does not exceed the threshold specified by the SLA. The throughput guarantee model allows a user to select optimal data transfer parameters that minimizes energy consumption while ensuring the achieved throughput does not fall below the throughput threshold specified in the SLA. We utilize a Matlab solver to obtain optimal transfer parameters based on the specified SLA.

After the offline analysis is performed, we store optimized kernel-level and application-level transfer parameters 
in our custom data structure for our dynamic online program to use.  Each data structure instance corresponds to the optimal parameter configuration of a cluster satisfying the SLA.  

\begin{algorithm}
\caption{Dynamic Energy Constraint Tuning}
  datasets = ClusterFiles() \\
  \For{Timeout} 
  {
     calculateDeltaEnergy()        \\
     calculateWeightedThroughput() \\
     calculateDeltaRtt()    \\
     $t = dataSize / Tavg$    \\
     $E_{pred} = P_{avg} * t$      \\
     calculateExternalNetworkPercentage()\\
     startWithLightExtNetworkLoadSurface()\\
     \tcc{If Energy Increased}
     \If {$(\Delta E_{last} + E_{pred} > (1+\beta) * pastE_{pred}) \parallel (\Delta E_{last} + E_{pred} > SLA)$} 
     {
          \tcc{External Network Load Increased}
          \If{$Ext_{Net} > (1 + \beta) * Ext_{ref_Net}$}
          {
                \tcc{Obtain Cluster/Surface with Higher External Network Load}
                $surface_{High\theta} \leftarrow NewOptParams$
          }
    }
    \tcc{Energy Decreased \& Ext Network Load Decreased}
    \ElseIf {$(refExt_{Net} < (1 - \alpha) * Ext_{NetLast})$}
    {
        $surface_{Low\theta} \leftarrow NewOptParams$
    }
} 
\end{algorithm}

\subsection{Online Dynamic Energy Constraint Tuning }

Cross-LayerHLA employs a dynamic online energy constraint tuning algorithm shown in Algorithm 1 derived from the offline energy constraint optimization model. Based on the energy constraint SLA, 
it periodically monitors the instantaneous power consumption at specified regular time intervals. If the instantaneous power consumption exceeds the threshold specified in the SLA, it obtains new optimal parameters within the confidence range/sampling region by obtaining a new energy polynomial surface that is closest to the current measured external network load and energy consumption range. If the measured instantaneous power consumption decreased from the last interval check, it obtains the closest energy polynomial surface and retrieves the associated optimal data transfer parameters. This is done approximately three times as continuously changing parameters are expensive.

\begin{algorithm}
\caption{Dynamic Throughput Tuning}
  datasets = ClusterFiles() \\
  \For{Timeout}
   {
     calculateWeightedThroughput() \\
     calculateDeltaRtt()    \\
     calculateExternalNetworkPercentage()\\
     startWithLightExtNetworkLoadSurface()\\
     \tcc{If Throughput Decreased}
     \If { $T_{avg} < (1-\alpha) * T_{Last} \parallel (T_{avg} < SLA) $  } 
     {
       \tcc{External Network Load Increased}
        \If{$Ext_{Net} > (1 + \beta) * refExtNet$}
        {
             \tcc{Obtain Cluster/Surface with Higher External Network Load}
              $surface_{High\theta} \leftarrow NewOptParams$
        }
    }    
    \Else { 
        \tcc{Throughput Increased}
        \If {
                \tcc{External Network Load decreased}
                 $refExt_{Net} < (1 - \alpha) * Ext_{NetLast}$
              } {\tcc{Obtain Cluster/Surface with Lower External Network Load}
           $surface_{Low\theta} \leftarrow NewOptParams$}
    }
} 
\end{algorithm}

\begin{figure*}[t]
    \begin{centering}
	\begin{subfigure}[t]{0.31\textwidth}
    	\centering
    	\caption{Chameleon Throughput (Mbps)}
        \includegraphics[keepaspectratio=true,width=58mm]{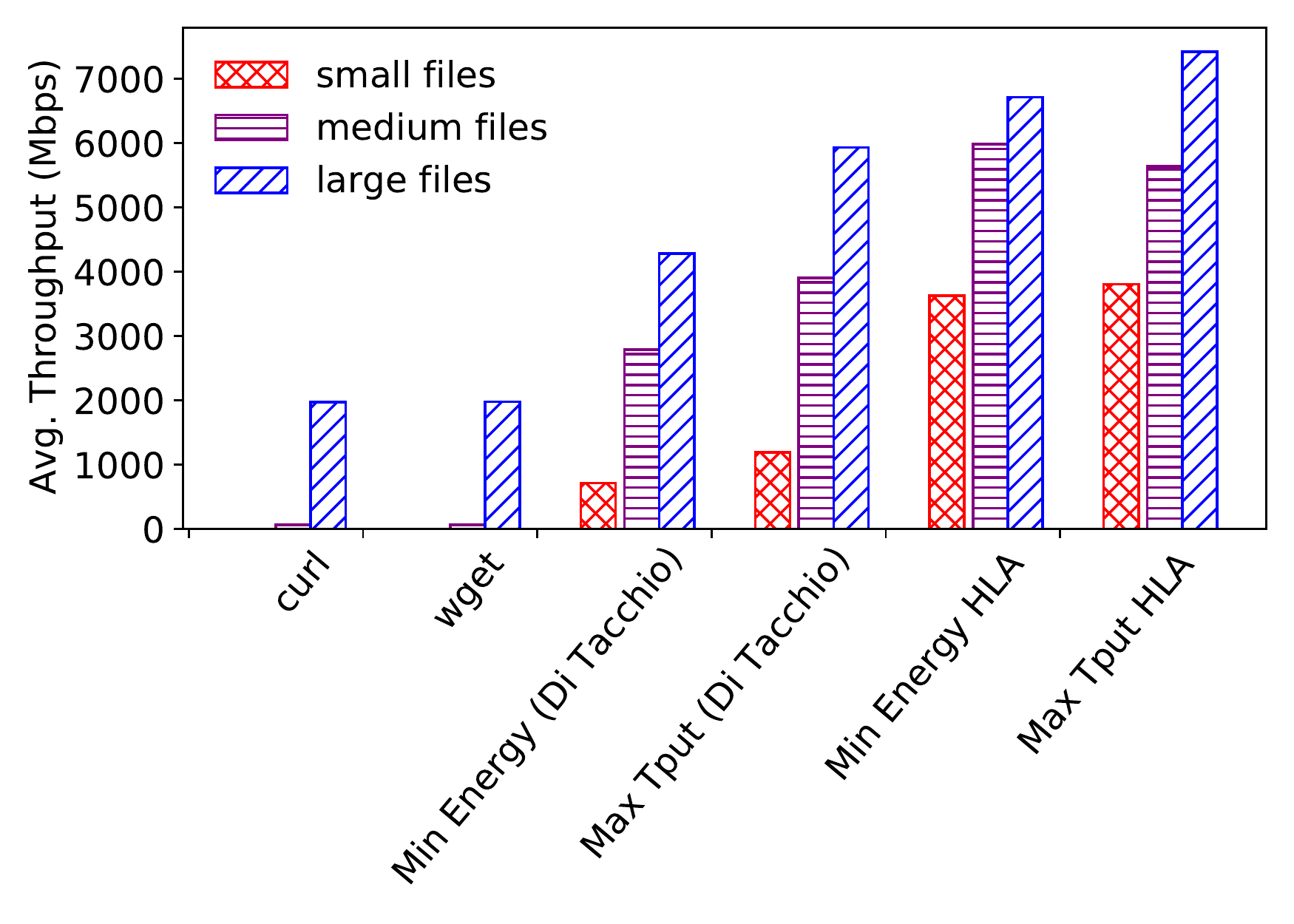}
        
        \vspace{-2mm}
    \end{subfigure}
    \begin{subfigure}[t]{0.31\linewidth}
    	\centering
    	\caption{CloudLab Throughput (Mbps)}
        \includegraphics[keepaspectratio=true,width=58mm]{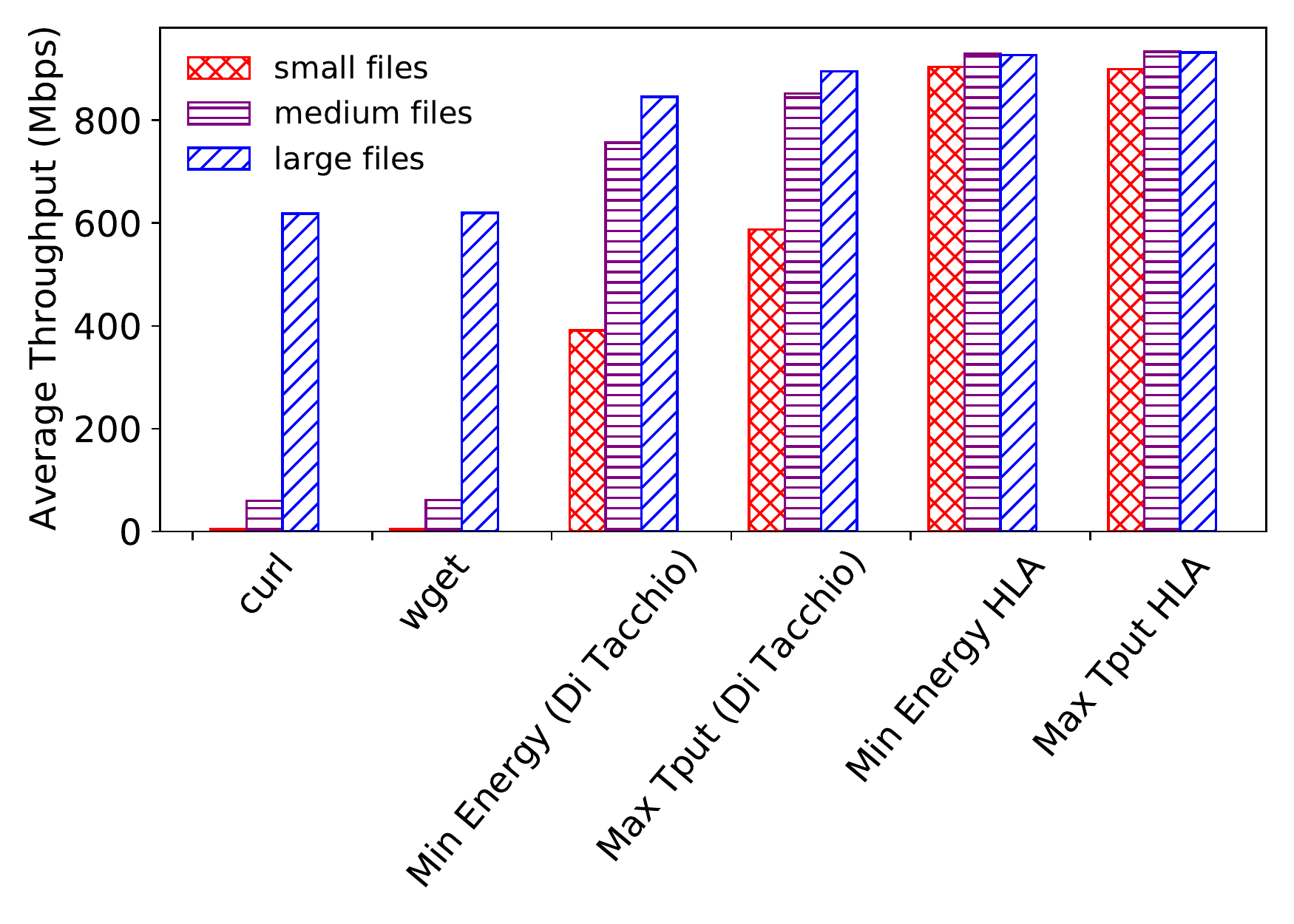}
        \vspace{-2mm}
    \end{subfigure}
    \begin{subfigure}[t]{0.31\textwidth}
    	\centering
    	\caption{Inter-Cloud Throughput (Mbps)}
        \includegraphics[keepaspectratio=true,width=58mm]{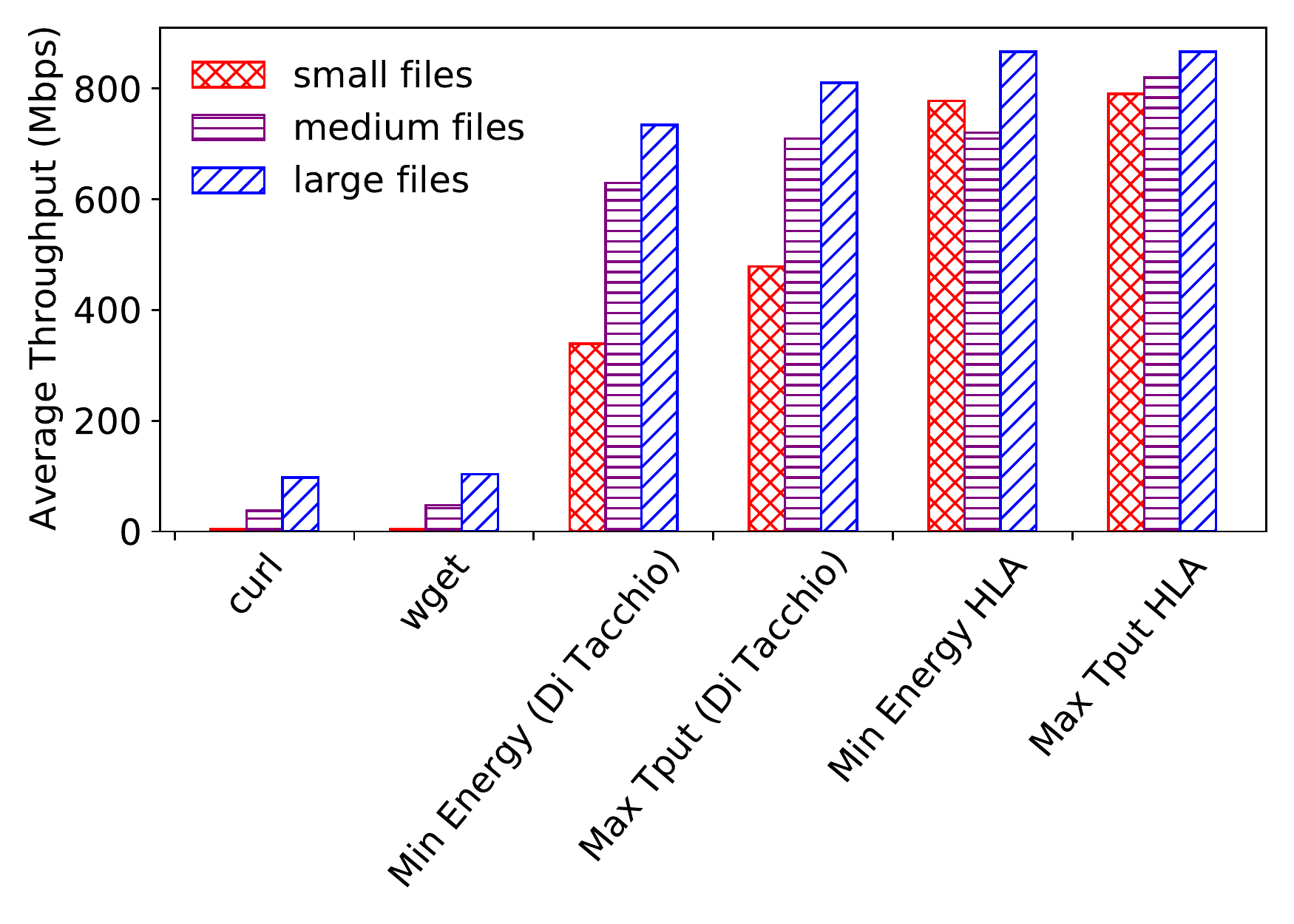}
        \vspace{-2mm}
    \end{subfigure}
	\begin{subfigure}[t]{0.31\textwidth}
    	\centering
    	\caption{Chameleon Client Energy (Joules)}
        \includegraphics[keepaspectratio=true,width=58mm]{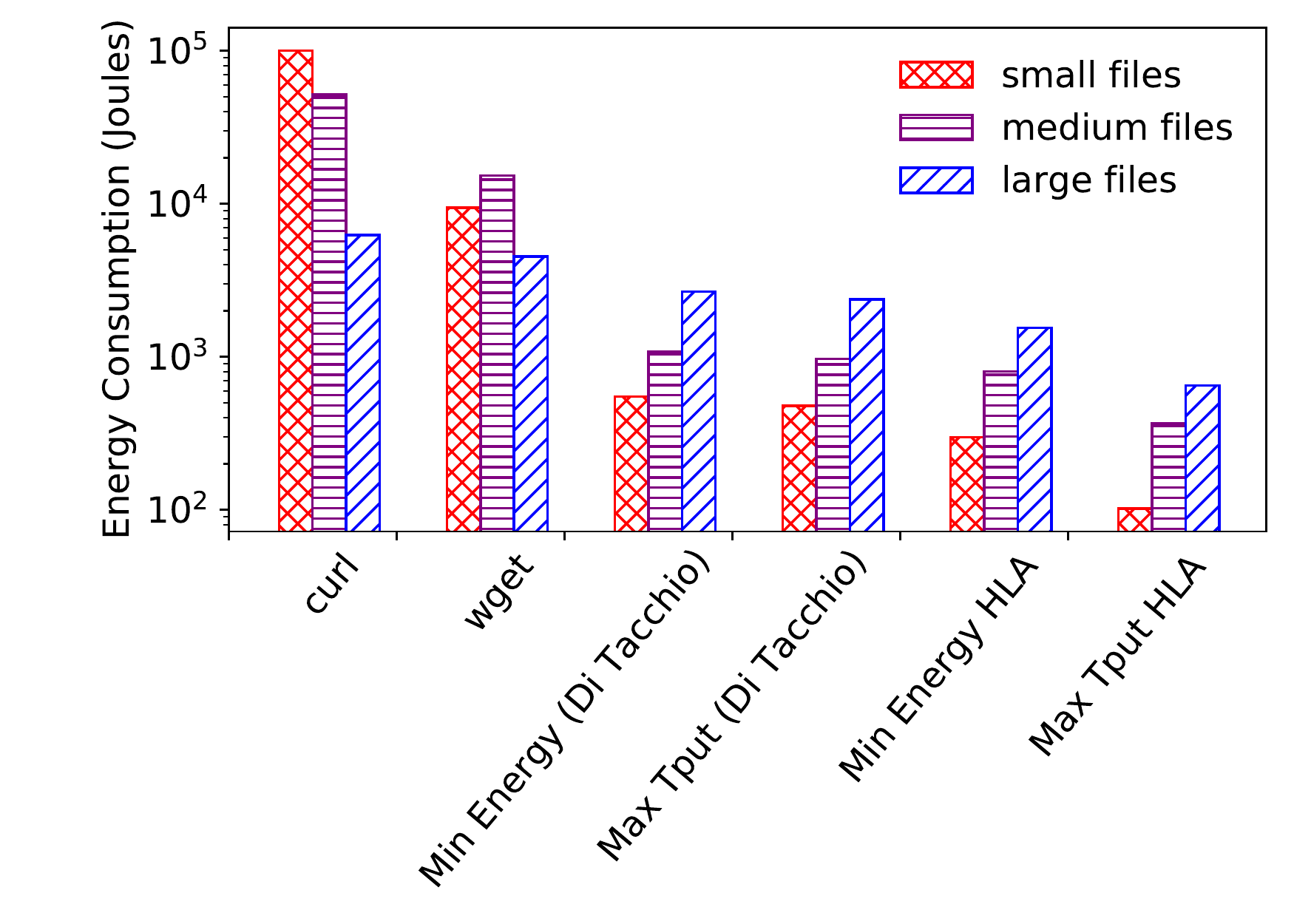}
        \vspace{-2mm}
    \end{subfigure}
 	\begin{subfigure}[t]{0.31\textwidth}
    	\centering
    	\caption{CloudLab Client Energy (Joules) }
        \includegraphics[keepaspectratio=true,width=58mm]{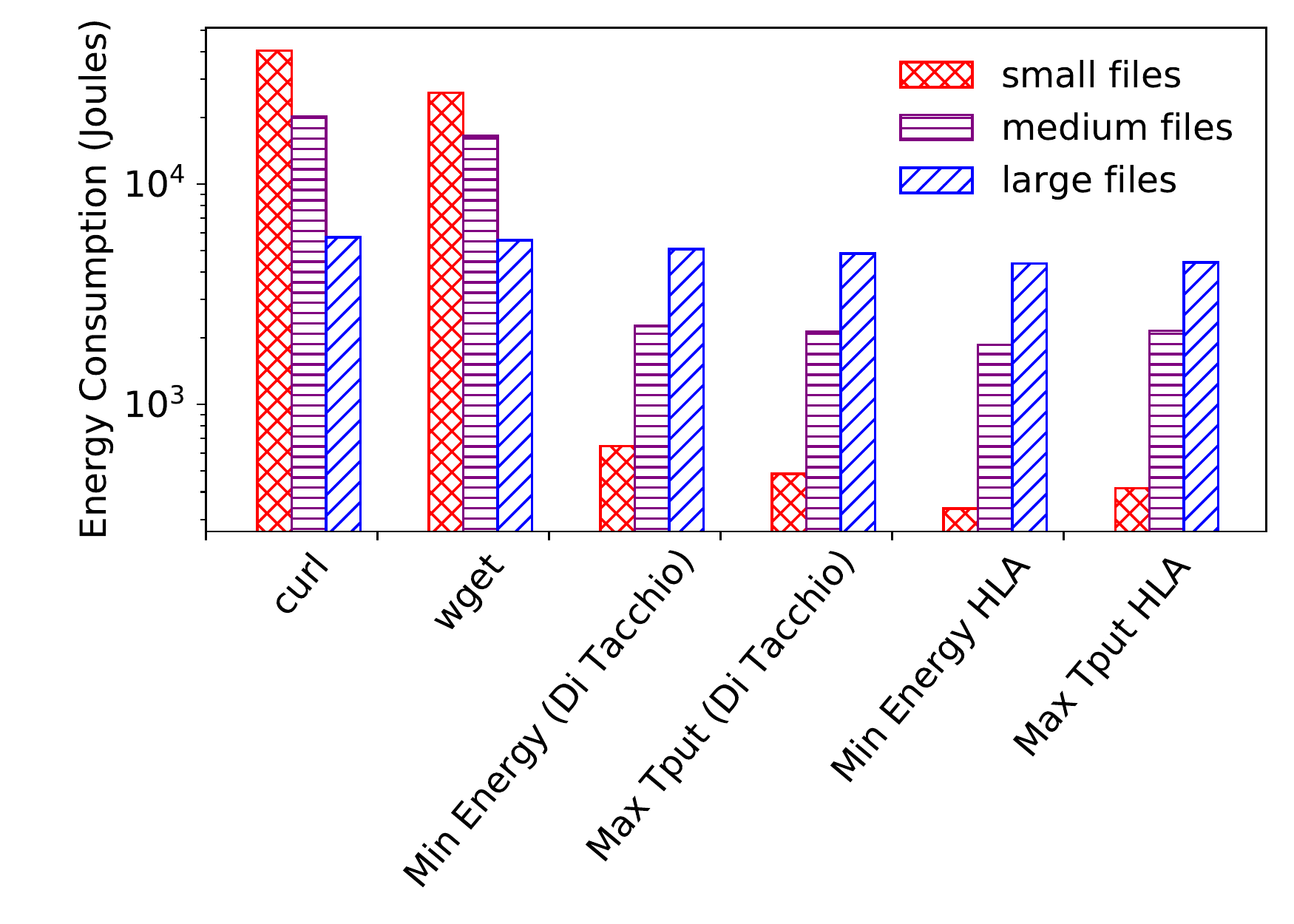}
        \vspace{-2mm}
    \end{subfigure}
    \begin{subfigure}[t]{0.31\textwidth}
    	\centering
    	\caption{Inter-Cloud Client Energy (Joules)}
        \includegraphics[keepaspectratio=true,width=58mm]{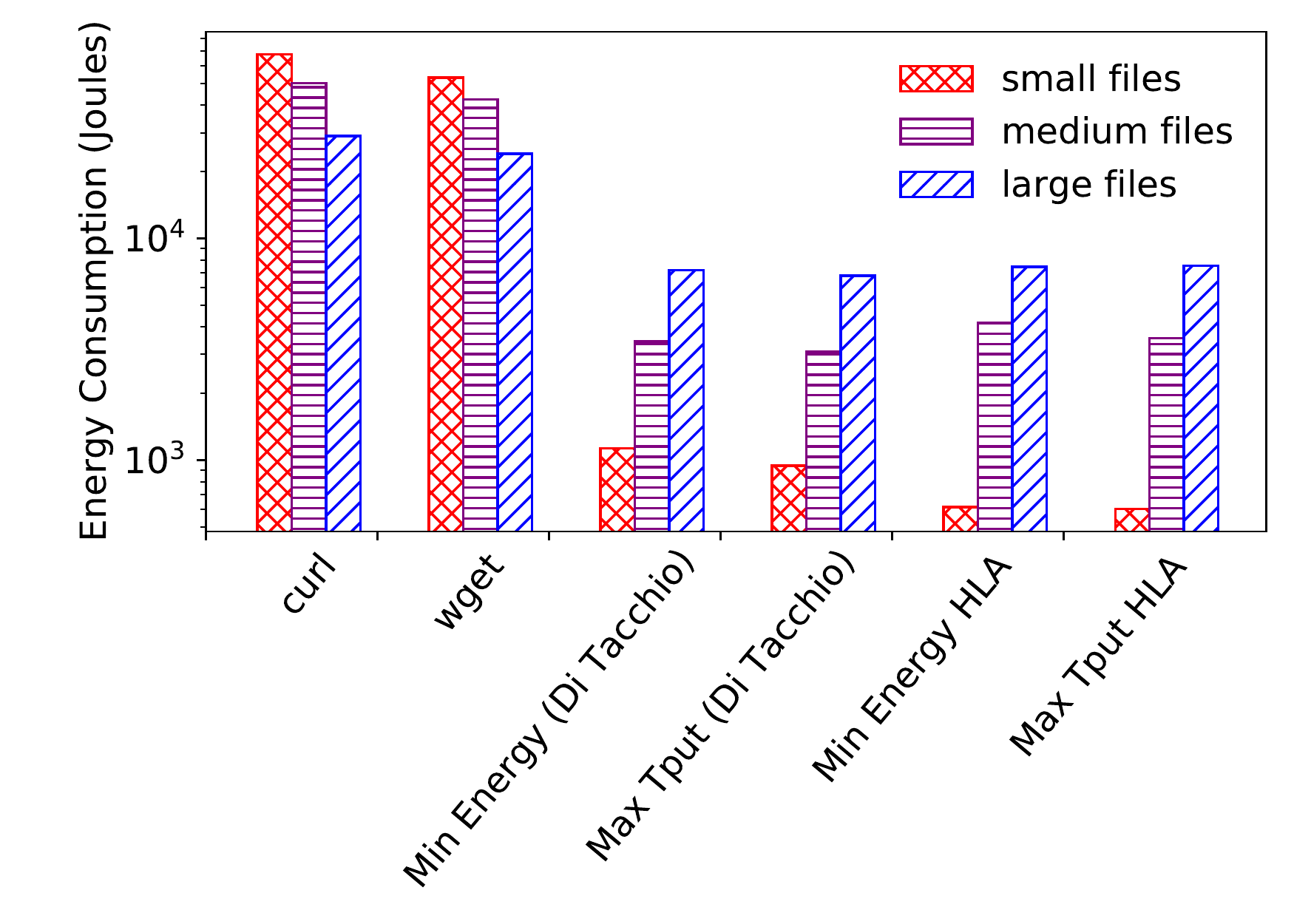}
        \vspace{-2mm}
    \end{subfigure}
     \caption{Achieved throughput (Mbps) and energy consumption (Joules) over 3 diverse testbeds}
     \vspace{-4mm}
     \label{fig:throughputresults}
     \end{centering}
 \end{figure*}

\subsection{Online Dynamic Throughput Guarantee Constraint Tuning }
Cross-LayerHLA employs a dynamic online throughput constraint tuning algorithm shown in Algorithm 2 derived from the offline throughput constraint optimization model. In this algorithm, it periodically monitors the instantaneous throughput at specified regular time intervals. If the instantaneous throughput falls below the threshold specified in the SLA, it obtains new optimal parameters within the confidence range/sampling region by obtaining a new polynomial throughput surface that is closest to the current measured external network load. If the measured instantaneous throughput increased or decreased from the last interval check, it obtains the closest throughput polynomial surface and retrieves the associated optimal data transfer parameters for the associated external network load. Obtaining optimal parameters from the polynomial throughput surfaces is done at a maximum of three times since it is expensive to modify the data transfer parameters. Afterward, if necessary, data transfer parameters are adjusted heuristically.

\subsection{Testing Online Dynamic Algorithms }
In order to fairly compare the efficiency of our dynamic energy constraint algorithm and throughput guarantee algorithm with other transfer tools/solutions, we utilize the extreme use cases. To achieve this, we use an SLA policy that informs our dynamic energy constraint algorithm to transfer data with the least amount of energy/power consumption by utilizing joint optimal kernel-level and application-level parameters. We call this SLA policy the minimum energy consumption SLA. To test our throughput guarantee algorithm, we use an SLA policy that enforces our algorithm to transfer data with the maximum throughput rate achievable, utilizing joint optimal kernel-level and application-level parameters.

\begin{table}[!t]
\centering
\label{t_sim}
\begin{tabular}{|l|l|l|l|l|}
\hline
\multicolumn{1}{|l|}{\textbf{Testbed}} & \multicolumn{1}{l|}{\textbf{Bandwidth}} & \multicolumn{1}{l|}{\textbf{RTT}} & \multicolumn{1}{l|}{\textbf{BDP}} &
\multicolumn{1}{l|}{\textbf{CPU Architecture}} \\ 
\hline
Chameleon & 10 Gbps & 32 ms & 40 MB & \makecell{Haswell (server) \\ Haswell (client)} \\  
\hline
CloudLab & 1 Gbps & 36 ms & 4.5 MB & \makecell{Haswell (server) \\ Broadwell (client)} \\
\hline
Inter-Cloud & 1 Gbps & 48 ms & 6 MB & \makecell{Haswell (server) \\ Bloomfield (client)} \\
\hline
\end{tabular}
\caption{Characteristics of testbeds}
\end{table}

\begin{table}[!t]
\centering
\label{t_sim}
\begin{tabular}{|l|l|l|l|l|}
\hline
\multicolumn{1}{|l|}{\textbf{Dataset}} & \multicolumn{1}{l|}{\textbf{Num Files}} & \multicolumn{1}{l|}{\textbf{Total Size}} & \multicolumn{1}{l|}{\textbf{Avg. File Size}} &
\multicolumn{1}{l|}{\textbf{Std. Dev.}} \\ 
\hline
Small & 20,000 & 1.94 GB & 101.92 KB & \makecell{29.06 KB} \\  
\hline
Medium & 5,000 & 11.70 GB & 2.40 MB & \makecell{0.27 MB} \\
\hline
Large & 128  & 27.85 GB & 222.78 MB & \makecell{15.19 MB} \\
\hline
\end{tabular}
\caption{Dataset Characteristics}
\end{table}
 
\section{Experimental Evaluation}

We collected experimental data by performing data transfers on three different wide-area network testbeds over a one week period.  
Our testbeds include (1) Chameleon Cloud, server located at the University of Chicago and client located at the Texas Advanced Computing Center; (2) CloudLab, server located at the University of Wisconsin and client located at the University of Utah; (3) Inter-Cloud, server located at the Texas Advanced Computing Center (part of Chameleon Cloud) and client located at the University of Wisconsin (part of CloudLab). Both the Chameleon nodes run on a Dell PowerEdge R630 containing 24 CPU cores distributed in dual-socket Intel Xeon E5-2670 v3 "Haswell" processors, each containing 12 cores and 128 GiB of RAM ~\cite{keahey2020lessons}. The client within the CloudLab architecture runs on an HPE ProLiant XL170r server containing 10 CPU cores plus hyper-threading distributed in an Intel E5-2640v4 "Broadwell" processor containing 64 GiB of RAM. The server within the CloudLab testbed as well as the client within the inter-cloud testbed run on Cisco's UCS SFF 220 M4 and UCS LFF 240 M4, respectively. Both contain 2 Intel E5-2630 "Haswell" processors, each containing eight cores plus hyper-threading and 128 GiB of RAM ~\cite{Duplyakin+:ATC19}.  The server within the Inter-Cloud testbed shares the same specifications as the Chameleon Cloud server. A specification overview is provided in table I. Experimental data transfers were performed during both peak hours and non-peak hours utilizing three diverse datasets containing different characteristics. These include: (1) the small file size dataset consisting of 20,000 HTML files derived from the common crawl project~\cite{CommonCrawl2020}; (2) the medium file size dataset consisting of 5,000 image files derived from Flickr~\cite{YahooFlickr}; (3) the large file size dataset consisting of 128 video files from Jiku~\cite{JikuVideoDataSet2020}. Complete dataset characteristics are specified in table II. We collected and stored data transfer meta-information on our log servers. We analyzed, grouped, and performed optimizations on our historical log data based on the target SLAs.  \par On all of our testbeds we measure the client's power consumption using Intel's Running Average Power Limit (RAPL) which uses a software model to accurately estimate power consumption based on hardware performance counters and I/O models. David et al.~\cite{5599016} and H\"ahnel et al.~\cite{hahnel2012measuring} highlighted RAPL's precision in measuring both memory and CPU power consumption. To distinguish data transfer power consumption from total system power consumption we subtracted the system baseline power consumption from the total power readings. In addition, to mitigate the disk-to-disk and I/O-to-disk transfer latency, we performed memory-to-memory data transfers. \par
To the best of our knowledge there are three related state-of-the-art solutions in this area. These include the algorithms proposed by Alan et al.~\cite{alan2015energy}, Di Tacchio et al.~\cite{di2019cross} and Nine et al.~\cite{nine2018greendataflow}. Alan et al.~\cite{alan2015energy} implemented energy-aware algorithms to optimize HTTP data transfers and reduce power consumption based on heuristics that lacked real-time tuning, adversely affecting throughput performance and end-system energy consumption during fluctuating network conditions. Di Tacchio et al. developed real-time tuning heuristics to optimize throughput and minimize energy consumption by tuning both application-level data transfer parameters and kernel-level parameters during HTTP transfers. However, at times, convergence to near optimal parameters was slow causing data transfers to utilize suboptimal parameter combinations for extended time periods during the tuning process. Slow convergence adversely affects throughput performance and increases end-system energy consumption. Over-estimating data transfer parameters and over- provisioning compute resources can increase  energy cost. Under-estimating data transfer parameters and under-provisioning compute resources can reduce throughput performance. 
Nine et al.~\cite{nine2018greendataflow} proposed dynamic tuning algorithms for GridFTP data transfers utilizing offline historical log analysis and online tuning mechanisms; however, they did not consider jointly scaling pertinent kernel-level parameters with data transfer application-level parameters. Resources such as the number of active cores and CPU frequency determine the number of Instructions per Second (IPS) that can be executed and affects throughput performance and end-system energy consumption. Dynamically tuning application-level parameters without tuning kernel-level parameters can 
adversely affect throughput performance and end-system energy consumption.\par 
Since we developed and implemented algorithms to dynamically tune HTTP data transfer parameters we compared our algorithms to those of Di Tacchio et al.~\cite{di2019cross}. 
Furthermore, we compared our algorithms to two common data transfer baseline tools: 1.) curl, an open source tool used to transfer data and 2.) wget, a free command line tool used to retrieve files from the web. 
For fair comparison between all algorithms and baseline tools we utilized the datasets specified in table II when performing experimental data transfers. In addition, since the baseline tools did not support service-level agreements (SLAs), we set the SLAs of our models/algorithms to two diametrical 
cases: (1) maximum achievable throughput (Max Tput HLA) and (2) minimum achievable energy consumption (Min Energy HLA).\par 
Figure 1 displays a comparison of throughput performance and energy consumption across three diverse testbeds: 1.) Chameleon Cloud, 2.) CloudLab and 3.) Inter-Cloud. As anticipated, curl and wget produced subpar results 
across all testbeds for all data transfers due to the absence of parameter optimization. Di Tacchio et al. algorithms performed better than all the baseline tools. However, executing  dataset transfers utilizing Di Tacchio et al. algorithms (Max Tput (Di Tacchio) and Min Energy (Di Tacchio) on the large Bandwidth-Delay-Product (BDP) testbed (Chameleon Cloud) illustrated a few  shortcomings: i) Heuristically tuning transfer parameters without using regression analysis or offline estimation techniques based on past data transfer history may cause slow convergence to optimal parameter values. This may be further exacerbated by fluctuating external network conditions which may cause the heuristic to overestimate application-level transfer parameters and compute resources such as the number of CPU cores. Additionally, dynamic external network conditions can cause the heuristic to underestimate application-level data transfer parameters and compute resources. ii) The cost of over-estimating or under-estimating parameter values and compute resources is expensive. Overestimating  parameter values and compute resources can increase energy consumption. Conversely, underestimating parameter values and compute resources can degrade throughput performance. 
Utilizing machine learning techniques and interpolation on past data transfer history logs allows our algorithms to accurately estimate near optimal data transfer parameters for a given SLA based on the current network conditions. This causes our algorithms to converge faster to optimal parameters. This is clearly observed in figure 1(a) and 1(d) which compares achieved throughput performance and data transfer energy across all algorithms and datasets. 
Max Tput HLA outperformed Max Tput (Di Tacchio)  across all datasets increasing throughput of HTML data transfers by 69\%, image data transfers by 31\% and video data transfers by 20\%. Furthermore, Max Tput HLA achieved the lowest power consumption due to reduced data transfer time. Extended data transfer time increases both static and non-static power consumption. Additionally, Min Energy HLA algorithm decreased energy consumption by an additional 46\% for HTML data transfers, 26\% for image data transfers and 42\% for video data transfers, with respect to Min Energy (Di Tacchio). Further throughput improvement was observed when we compared Max Tput HLA algorithms against the baseline tools. Max Tput HLA improved throughput performance for both HTML and image data transfers by up to 99\% when compared to curl and wget. For video data transfers, Max Tput HLA increased throughput by 73\%. Furthermore, Min Energy HLA decreased energy consumption by an additional 99\% for HTML data transfers, 98\% for image data transfers and 75\% for video data transfers. 
Moreover, for all algorithms we enhanced throughput performance and minimized end-system energy consumption utilizing approximately 67\% of the available CPU cores.\par 
Data transfers executed on the CloudLab network, a lower BDP network, demonstrated that Max Tput HLA outperformed Max Tput (Di Tacchio) by up to 35\% for HTML transfers. Max Tput HLA further increased throughput by 14\% for image data transfers and 8\% for video transfers. For lower BDP networks, both Max Tput (Di Tacchio) and Min Energy (Di Tacchio) were able to converge faster to optimal parameters than on larger BDP networks. Nevertheless, the heuristic tuning mechanism has the propensity to either overestimate or underestimate data transfer parameters and compute resources when network conditions fluctuate. On the otherhand, Min Energy HLA decreased energy consumption by an additional 48\% for HTML data transfers, 18\% for image data transfers and 9\% for video transfers compared to Min Energy (Di Tacchio). With respect to the baseline tools, Max Tput HLA further increased throughput by up to 99\% for HTML transfers, 94\% for image transfer and up to 34\% for video transfers. Furthermore, Min Energy HLA decreased energy consumption by an additional 99\% for HTML transfers and up to 24\% for video transfers. 
Experiments performed on the Inter-Cloud network demonstrated that the Max Tput HLA algorithm increased throughput by an additional 40\% for HTML data transfer, 13\% for image data transfers and 8\% for video data transfers with respect to Max Tput (Di Tacchio). Furthermore, Min Energy HLA decreased energy consumption by an additional 36\% for HTML data transfers, but slightly increased energy consumption for image and video transfers. Compared to curl, Max Tput HLA increased throughput for HTML data transfers by  99\%, 95\% for image data transfers and 89\% for video data transfers. In addition, Min Energy HLA decreased energy consumption by an additional 99\% for HTML data transfers, 92\% for image data transfer and 74\% for video data transfers. Compared to wget, Max Tput HLA increased throughput by 99\% for HTML data transfers, 95\% for image data transfers and  89\% for video data transfers. Furthermore, Min Energy HLA  decreased energy consumption by an additional 99\% for HTML data transfers, 93\% for image data transfers and 88\% for video data transfers. Additionally, for all algorithms we enhanced performance by approximately using only 80\% of the available CPU cores.

\par
\section{Conclusion}
In this paper, we have introduced a cross-layer optimization framework that combines offline analysis with adaptive online tuning to minimize end-system energy consumption and maximize data transfer throughput performance. We presented novel algorithms that dynamically tune both application-level and kernel-level parameters based on historical log analysis and current network conditions to meet the requirements set by the service-level agreements (SLAs). Our detailed experimental analysis and results show that our proposed Cross-LayerHLA algorithms outperforms the state-of-the-art solutions in this area, reducing energy consumption considerably while increasing data transfer throughput.

\section*{Acknowledgements}
This project is in part sponsored by the National Science Foundation (NSF) under award numbers 2007829, 1724898 and 1842054. We also would like to thank the Chameleon Cloud and CloudLab for letting us use their resources in our experiments.

\bibliographystyle{acm}
\bibliography{references}
\end{document}